\documentclass[twocolumn,english]{article}
\usepackage{lmodern}
\usepackage[T1]{fontenc}
\usepackage[latin9]{inputenc}
\usepackage[a4paper]{geometry}
\usepackage{float}
\usepackage{amsmath}
\usepackage{amssymb}
\usepackage{graphicx}
\usepackage{esint}

\makeatletter

\providecommand{\tabularnewline}{\\}

\numberwithin{equation}{section}
\numberwithin{figure}{section}
\newcommand{\lyxaddress}[1]{
\par {\raggedright #1
\vspace{1.4em}
\noindent\par}
}

\makeatother

\usepackage{babel}
\begin{document}

\title{Application of the variations calculus on energy spectra of shallow
hydrogen-like impurities and excitons in GaAs/AlGaAs quantum wells}

\author{Gubanov A.\textsuperscript{1,2}, Glinskii G. F.\textsuperscript{1}}

\date{(Dated 29 April 2009)}

\maketitle

\lyxaddress{\textsuperscript{1}Saint Petersburg Electrotechnical Univeristy
``LETI'', Saint Petersburg, 197376, Russian Federation}

\lyxaddress{\textsuperscript{2}Corresponding author: \includegraphics[scale=0.33]{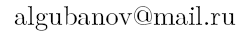}}
\begin{abstract}
A variational calculus method is applied for hydrogen-like impurities
states and exciton states in AlGaAs/GaAs quantum wells within the
effective masses approach. The hamiltonian is separated into radial
and transversal parts to be considered and optimized using trial wavefunctions.
The results satifsy the well-known cases of three-dimentional and
two-dimentional hydrogen. The donor and exciton bound energy is shown
to decrease while widthening the quantum well. The non-symmetrical
donor position also decreases the bound energy.
\end{abstract}

\section{Introduction}

Donor and acceptor energy spectra determination in the quantum well
(QW) structures has a consistent interest combining theoretical accent
and practical issues. Impurity ionisation energy defines the spatial
charge distribution in heterostructures, while the excitonic spectra
governs the optical properties near the QW band gap region\cite{Glinskii}. 

This paper focuses on shallow hydrogen-like impurities and excitons
in single QW structures of GaAs/AlGaAs in assumption of the effective
mass approach. The developed variations calculus method minimizes
lateral part of the hamiltonian, in contrast to the typical calculation
based on the total hamiltonian functional minimization \cite{Parr,Ponomarev}.
The wavefunction is assumed as a product of 2D trial and a function
of spatial coordinate directed normally to the QW plane. In this framework
the problem for hydrogen-like impurities is reduced to 1D Shroedinger
equation solution, and 2D solution for exciton calculation.

The obtained diffential equations were solved numerically using the
variations calculus and applied for $1s$, $2p_{0}$ and $2p_{\pm}$
states of hydrogen-like impurity and exciton in GaAs/AlGaAs QW as
described in Section 2. Energy spectra dependencies on the QW width,
depth and the inpurity position were considered in Section 3.

\section{Calculation procedure}

\subsection{Donor states}

\subsubsection{Trial wavefunction choice }

Abrupt QW is placed normally to the $z$ direction. Thus the hamiltonian
describing the electron in the box QW potential $U\left(z\right)$
and the donor Coulombic field is
\[
\hat{H}=-\frac{\hbar^{2}}{2m_{\textrm{eff}}}\nabla^{2}+U\left(z\right)-\frac{e^{2}}{\epsilon\left|\mathbf{x}-\mathbf{x}_{b}\right|},
\]

where $\mathbf{x}_{b}=\left\{ 0,0,b\right\} $ aims the donor position.
Since now we operate dimentionless units. The axial symmetry is benefited
in a cylindrical coordinates $\left(\rho,\phi,z\right)$, where $\hat{H}$
becomes
\begin{equation}
\hat{H}=-\nabla^{2}+U\left(z\right)-\frac{2}{\sqrt{\rho^{2}+\left(z-b\right)^{2}}}.\label{eq:dHdim}
\end{equation}
The Shrödinger equation $\hat{H}\Psi=E\Psi$ determines a total wavefunction
$\Psi$. We consider this as a product of 2D trial and a function
of spatial coordinate across the QW

\[
\Psi\left(\rho,\phi,a,z\right)=F\left(\rho,\phi,a\right)\xi\left(z\right),
\]
where $a$ is a variation parameter, $F\left(\rho,\phi,a\right)$
notes an trial 2D part operating in a QW plane $\left(x,y\right)$
and being independent on $z$, while the last term $\xi\left(z\right)$
defines the wavefunction along $z$ axis. The 2D part is bounded at
the infinite range and symmetrical. 

\begin{table}[H]
\begin{tabular}{cc}
\hline 
State & 2D wavefunction part\tabularnewline
\hline 
$1s$ & $A_{\textrm{1s}}\exp\left(-\rho/a\right)$\tabularnewline
$2p_{0}$ & $A_{\textrm{2p}}\exp\left(-\rho/a\right)$\tabularnewline
$2p_{-}$ & $A_{\textrm{2p}-}\rho\cos\phi\exp\left(-\rho/a\right)$\tabularnewline
$2p_{+}$ & $A_{\textrm{2p+}}\rho\cos\phi\exp\left(-\rho/a\right)$\tabularnewline
\hline 
\end{tabular}

\caption{Trial radial parts of wavefunction }
\end{table}

The pre-exponential factor $A$ is chosen to satisfy the unite propability
of the particle presence in heterostructure
\begin{equation}
\intop_{0}^{+\infty}\intop_{-\pi}^{\pi}\left|F\left(\rho,\phi,a\right)\right|^{2}\rho d\rho d\phi=1.\label{eq:norm}
\end{equation}

\subsubsection{Shrödinger equation solution}

The Shrödinger equation $\hat{H}\Psi=E\Psi$ is multiplied with a
2D ansatz $F\left(\rho,\phi,a\right)$ and integrated over the QW
plane $\left(x,y\right)$. Using \eqref{eq:norm} condition one obtains
another operator equation for the longitudinal wavefunction part $\xi\left(z\right)$
\[
\hat{K}\left(a,z\right)\xi\left(z\right)=E\xi\left(z\right),
\]
where $\hat{K}$ is a mean hamiltonian value over the ansatz function
$F\left(\rho,\phi,a\right)$. For a $n$-th state, this one yields

\[
\hat{K}_{n}\left(a,z\right)=\intop_{0}^{+\infty}\intop_{-\pi}^{\pi}F_{n}\left(\rho,\phi,a\right)\hat{H}F_{n}\left(\rho,\phi,a\right)\rho d\rho d\phi=
\]
\[
=\intop_{0}^{+\infty}\intop_{-\pi}^{\pi}F_{n}\left(\rho,\phi,a\right)\left\{ -\left[\frac{1}{\rho}\nabla_{\rho}+\nabla_{\rho}^{2}+\frac{1}{\rho^{2}}\nabla_{\phi}^{2}+\nabla_{z}^{2}\right]+\right.
\]
\[
+\left.U\left(z\right)-\frac{2}{\sqrt{\rho^{2}+\left(z-b\right)^{2}}}\right\} F_{n}\left(\rho,\phi,a\right)\rho d\rho d\phi=
\]
\[
=-\nabla_{z}^{2}+U\left(z\right)+I_{n}\left(a,z\right).
\]

The problem is reduced to a 1D Shrödinger equation with additional
potential term $I_{n}\left(a,z\right)$. The subject of the variation
procedure related to $a$ parameter is to detect the minimal value
of $I_{n}\left(a,z\right)$ for any given $z$ by varying $a=a_{\textrm{min}}\left(z\right)$.
Finally, the effective operator is 

\[
\hat{K}_{n}\left(a,z\right)=-\nabla_{z}^{2}+U\left(z\right)+W_{n}\left(z\right),
\]

where $W_{n}\left(z\right)=I_{n}\left(a_{\textrm{min}}\left(z\right),z\right)$
and acts as additional effective potential. Its profile depends only
on state considered and does not involve the QW parameters. Figure
2.1 depicts $W_{n}\left(z\right)$ profile for $n=$ $1s$, $2p_{0}$
and $2p_{\pm}$ states. 

\begin{figure}
\includegraphics{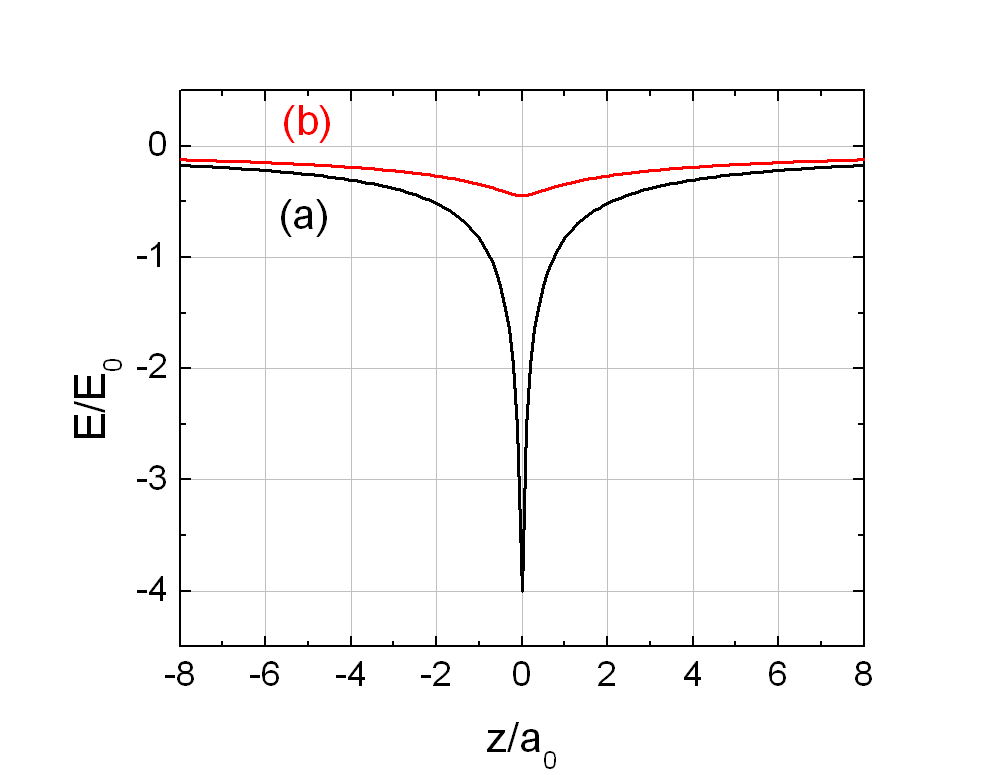}

\caption{Effective potential $W_{n}\left(z\right)$ profile for (a) $n=$ $1s$,
$2p_{0}$ and (b) $2p_{\pm}$ states.}
\end{figure}

1D Shrödinger equation
\[
\left[-\nabla_{z}^{2}+U\left(z\right)+W_{n}\left(z\right)\right]\xi\left(z\right)=E\xi\left(z\right)
\]
 was solved numerically in a routine approach \cite{Glinskii}. Here
we neglect the difference in lattice pediods and effective carrier
masses within the materials of the heterostructure. The strains are
also out of the scope due to the matched AlGaAs/GaAs material choice.

\subsection{Exciton states}

The hamiltonian for an excition problem operates with the electron
$z_{\textrm{e}}$ and hole $z_{\textrm{h}}$ coordinates and comprises
twice more degrees of freedom

\[
\hat{H}=-\frac{\hbar^{2}}{2m_{\textrm{e,eff}}}\nabla_{\textrm{e}}^{2}-\frac{\hbar^{2}}{2m_{h,\textrm{eff}}}\nabla_{\textrm{h}}^{2}+
\]
\[
+U_{\textrm{e}}\left(z_{\textrm{e}}\right)+U_{\textrm{h}}\left(z_{\textrm{h}}\right)-\frac{e^{2}}{\epsilon\left|\mathbf{x}_{\textrm{e}}-\mathbf{x}_{\textrm{h}}\right|}.
\]

After switching to the mass center and performing variational procedure
as in section 2.1.2, the problem reduces to 1D case regarding $z_{\textrm{e}}$
and $z_{\textrm{h}}$, while effective potential $W_{n}\left(z\right)$
becomes equal to the one of donor case

\[
\left[-\eta_{\textrm{e}}\nabla_{\textrm{e}}^{2}-\eta_{\textrm{h}}\nabla_{\textrm{h}}^{2}+U_{\textrm{e}}\left(z_{\textrm{e}}\right)+U_{\textrm{h}}\left(z_{\textrm{h}}\right)+W_{n}\left(z_{\textrm{e}}-z_{\textrm{h}}\right)\right]\xi=E\xi,
\]

where notation $\eta_{\textrm{e}}=\mu/m_{\text{e}}$, $\eta_{\textrm{h}}=\mu/m_{\text{h}}$
are used and $\mu=1/\left(m_{\text{e}}^{-1}+m_{\text{h}}^{-1}\right)$
is the dimensionless reduced exciton mass.

\section{Results and discussion}

To verify the variational procedure, the calculation approach was
applied to well-known 3D and 2D hydrogen problem. The well depth was
set to $U\left(z\right)=0$ and $U\left(z\right)=-\infty$ respectively,
to obtain $E_{0}$ and $E_{1}$ energies. The results in the Rydberg
units are presented in Table 2 and indicate appropiate accuracy. 

\noindent \begin{flushleft}
\begin{table}[H]
\begin{tabular}{lcccc}
\hline 
 & $E_{0}$/Ry & $E_{1}$/Ry & d$E_{0}$ & d$E_{1}$ \tabularnewline
\hline 
Our solution & \textendash{}1.083 & \textendash{}0.249  & 8.3 \% & 0.4 \%\tabularnewline
Exact solution & \textendash{}1.000 & \textendash{}0.250  & $-$ & $-$\tabularnewline
\hline 
\end{tabular}

\noindent \raggedright{}\caption{The variational procedure accuracy.}
\end{table}

\par\end{flushleft}

The bound energy dependecies of the QW width were considered in assumption
the donor was positioned in the AlGaAs/GaAs well center. As shown
in Fig. 3.1.-3.2. the decrease in the well width increases the binding
energy of the donor electron and exciton energy as well. This is attributed
to severe wavefunction concentration in the narrow QW \cite{Gubanov1,Gubanov2}. 

\begin{figure}
\includegraphics{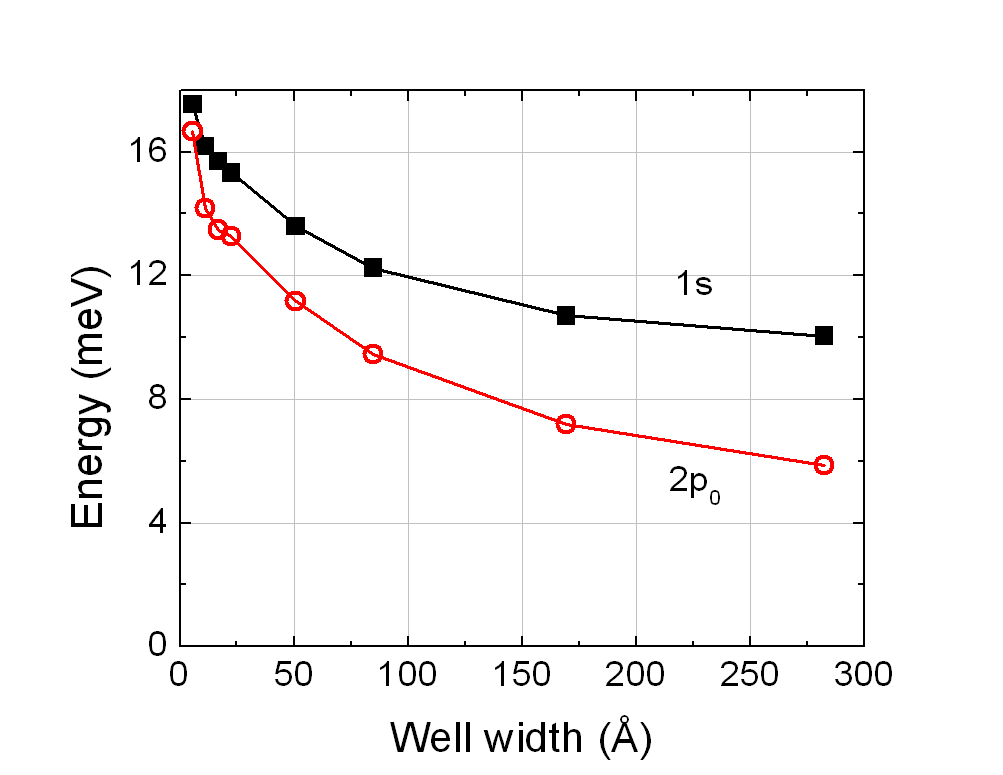}

\caption{Donor state bound energy for $n=$ $1s$, $2p_{0}$ vs QW width. }

\end{figure}

\begin{figure}
\includegraphics{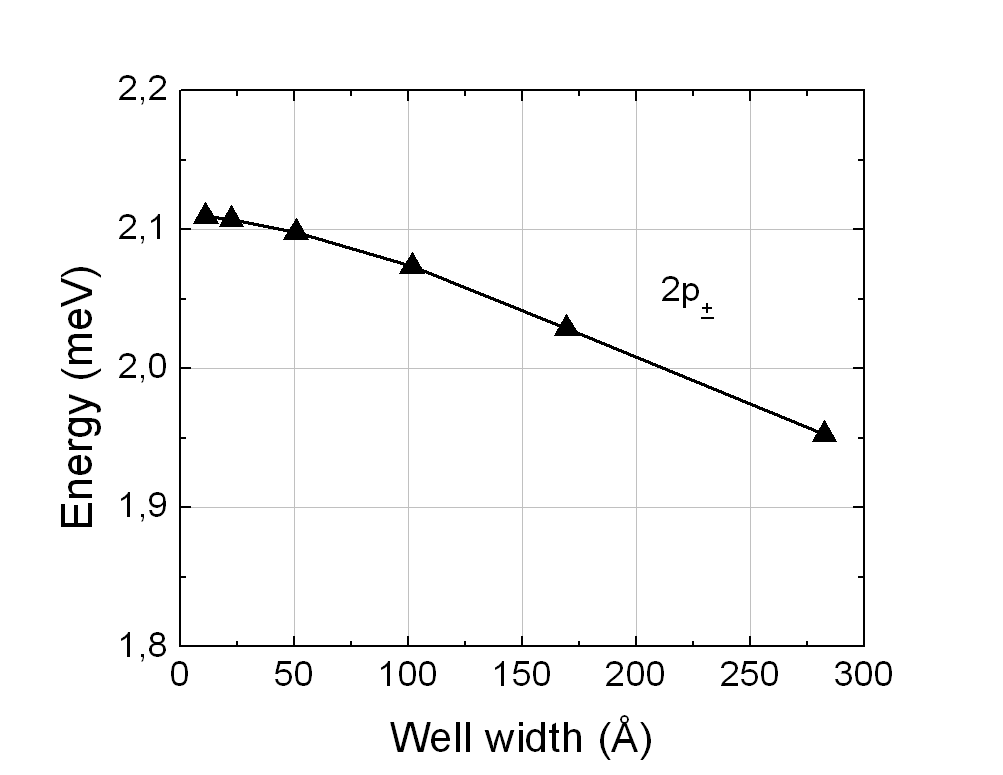}

\caption{Donor state bound energy $2p_{\pm}$ vs QW width.}
\end{figure}

Then we considered the donor to be shifted away the QW center. Figures
3.3. and 3.4. illustrate the results. The displacement decreases the
bound energy and provokes significant disturbance while reaching the
QW border. The intermixing of $1s$ and $2p_{0}$ states occurs near
30 Å position and may be interpreted as computational feature, so
that $1s$ should prevail the $2p_{0}$ after 30 Å as formerly. The
general decrease indicates the wavefunction diffusion from the QW
region. 

\begin{figure}[h]
\includegraphics{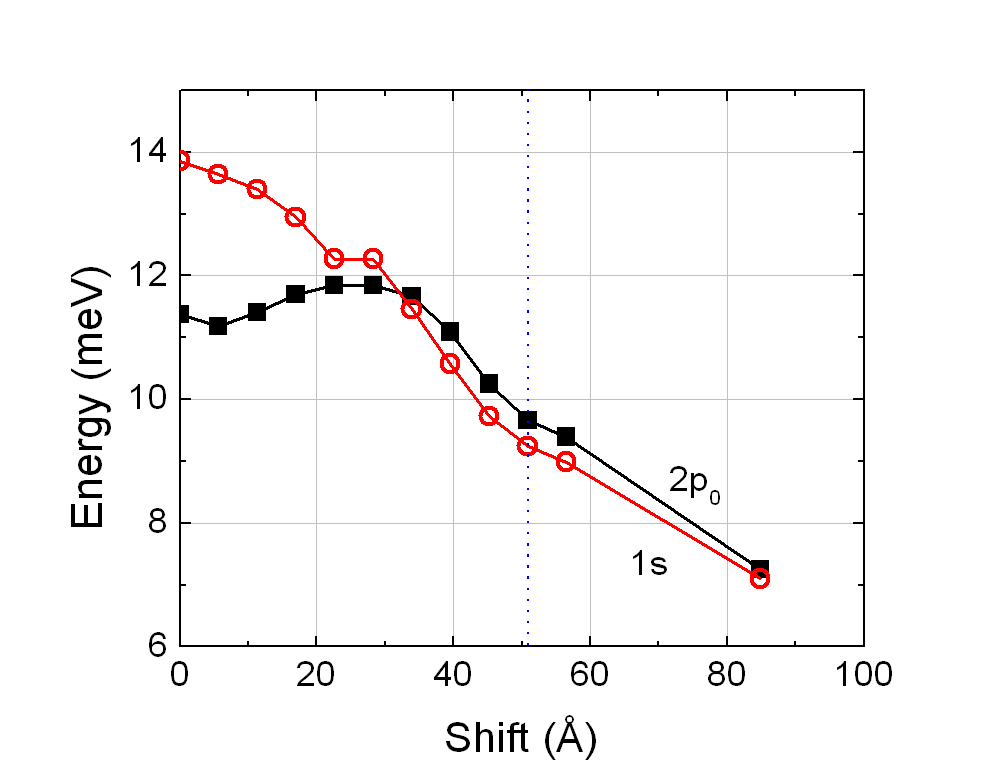}

\caption{Donor state bound energy for $n=$ $1s$, $2p_{0}$ vs donor displacement
against the center position in a 50 Å QW. The QW border is shown in
blue.}

\end{figure}

\begin{figure}[h]
\includegraphics{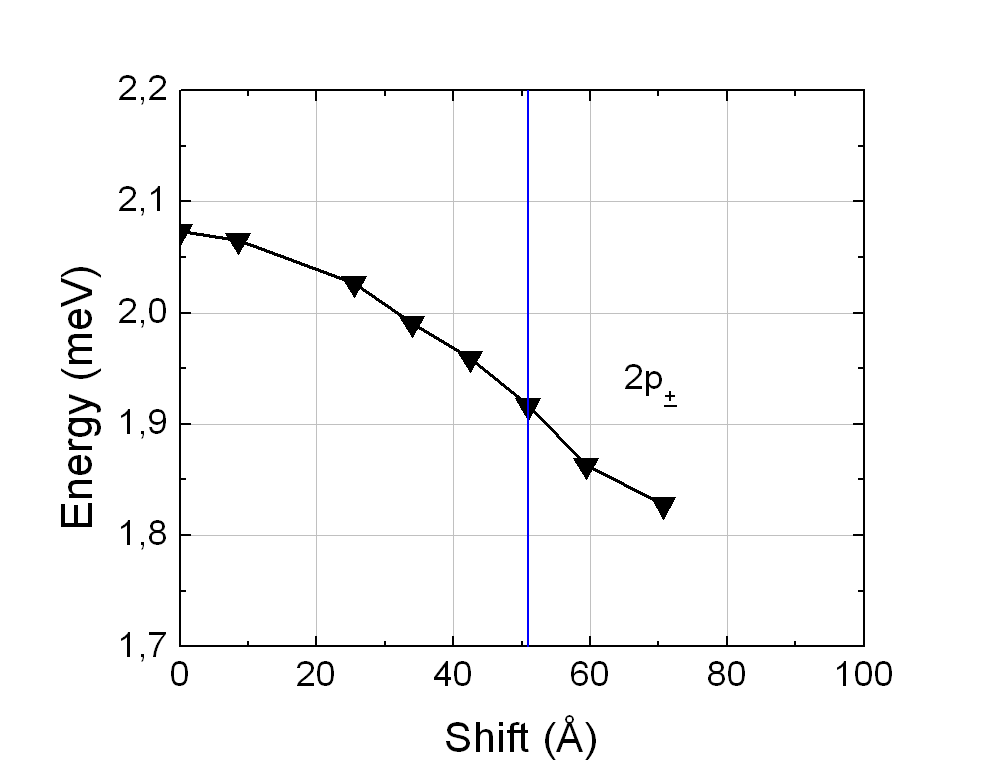}

\caption{Donor state bound energy $2p_{\pm}$ vs donor displacement against
the center position in a 50 Å QW. The QW border is shown in blue.}
\end{figure}

After that the exciton state was calculated in dependence on the QW
width. Similarly to the donor problem, the increase in width causes
the wavefunction diffusion and lowers the bound energy of exciton,
as shown in Figures 3.5. and 3.6. 

\begin{figure}[h]
\includegraphics{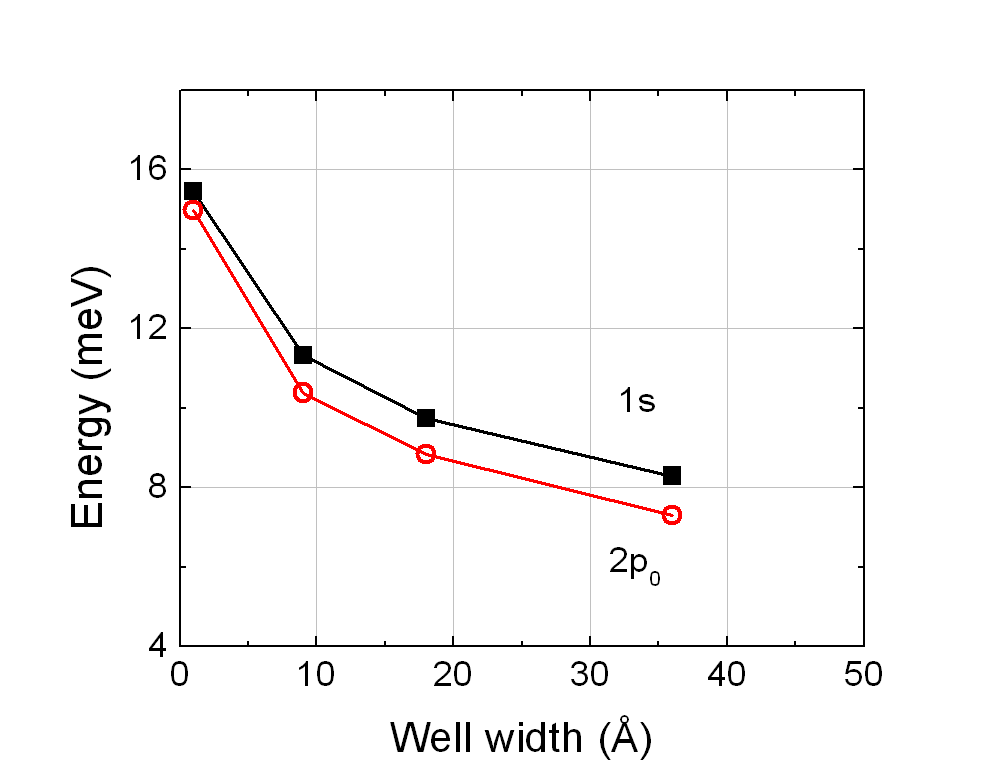}

\caption{Exciton bound energy for $n=$ $1s$, $2p_{0}$ vs QW width}

\end{figure}

\begin{figure}[h]
\includegraphics{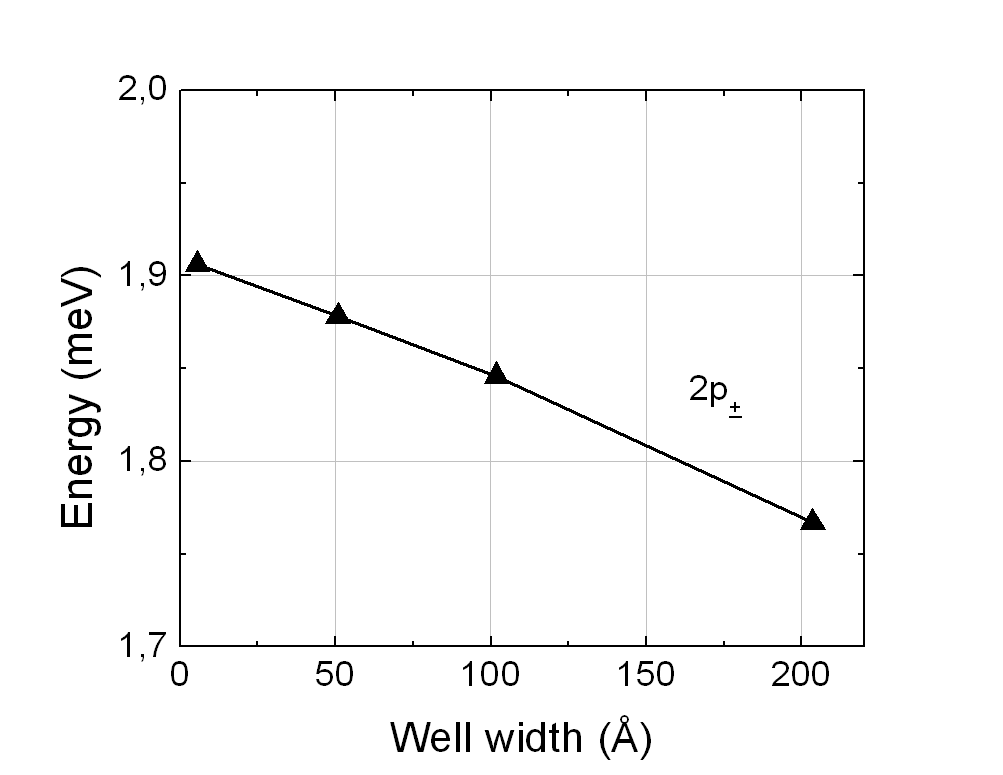}

\caption{Exciton bound energy $2p_{\pm}$ vs QW width}
\end{figure}

\section{Conclusions}

The developed variation method was applied for energy spectra calculations
for shallow hydrogen-like impurities and excitons in single GaAs/AlGaAs
quantum wells. The results for $1s$, $2p_{0}$ and $2p_{\pm}$ states
rather satisfy the theoretical results for typical cases. 

Shrinking the well width increases the binding energy of the donor
electron and exciton energy as well. The shift of impurity away of
QW center decreases the electron binding energy.

\section*{Acknowledgements}

This work has been conducted within the programme ``Scientific potential
development in universities (2009-2010)'' \#2503 and supported by
the Russian Fundamental Research Foundation (RFFI) grant \#08-02-00162.

\section*{Note}

This paper is a pre-print of the manuscript titled ``Small hydrogen-like
centers and excitons in structures with single quantum wells'' published
in \textquotedbl{}St. Petersburg State Polytechnical University Journal.
Physics and Mathematics\textquotedbl{} \textbf{3} (83) 2009 (ISSN
1994-2354) by the mentioned authors. 
\end{document}